\documentclass[aps,pra,amssymb,reprint,twocolumn,superscriptaddress,showpacs]{revtex4-1}

\usepackage[ansinew]{inputenc}
\usepackage{color}
\usepackage{graphicx}
\usepackage{amsmath}

\begin{document}


\title{Tracing transient charges in expanding clusters}

\author{Bernd Schütte}
\email{schuette@mbi-berlin.de}
\affiliation{Max-Born-Institut, Max-Born-Strasse 2A, 12489 Berlin, Germany}
\author{Marc J. J. Vrakking}
\affiliation{Max-Born-Institut, Max-Born-Strasse 2A, 12489 Berlin, Germany}
\author{Arnaud Rouz\'{e}e}
\affiliation{Max-Born-Institut, Max-Born-Strasse 2A, 12489 Berlin, Germany}

\date{\today}

\begin{abstract}
We study transient charges formed in methane clusters following ionization by intense near-infrared laser pulses. Cluster ionization by 400~fs ($I=1 \times 10^{14}$~W/cm$^2$) pulses is highly efficient, resulting in the observation of a dominant C$^{3+}$ ion contribution. The C$^{4+}$ ion yield is very small, but is strongly enhanced by applying a time-delayed weak near-infrared pulse. We conclude that most of the valence electrons are removed from their atoms during the laser-cluster interaction, and that electrons from the nanoplasma recombine with ions and populate Rydberg states when the cluster expands, leading to a \textit{decrease} of the average charge state of individual ions. Furthermore, we find clear bound-state signatures in the electron kinetic energy spectrum, which we attribute to Auger decay taking place in expanding clusters. Such nonradiative processes lead to an \textit{increase} of the final average ion charge state that is measured in experiments. Our results suggest that it is crucial to include both recombination and nonradiative decay processes for the understanding of recorded ion charge spectra. 
\end{abstract}

\maketitle


The ionization of clusters by intense laser pulses induces highly complex dynamics that take place on attosecond to nanosecond timescales and that involve a large number of interacting particles. Intense laser-cluster interactions commonly involve an ionization stage, the establishment of a nanoplasma and the subsequent expansion and break-up of this nanoplasma. Disentangling the different mechanisms is a very challenging task that requires sophisticated theoretical and experimental approaches. In spite of the large efforts that have been devoted to the study of laser-cluster interactions in the past 20 years (see e.g.~\cite{ditmire96,krainov02,saalmann06,fennel10}), the understanding of both the ionization and relaxation dynamics taking place during the cluster expansion are still far from being complete. For instance, up to now, experiments have only provided limited information about the relative importance of direct laser-induced (multi-photon and tunneling) ionization and electron-impact ionization by laser-driven electron-atom/ion collisions
                                                                         
The development and application of pump-probe techniques promises novel insights into the relevant processes. For instance, the generation of seed electrons in a cluster by an extreme-ultraviolet (XUV) pulse allows the time-resolved investigation of strong-field processes induced by near-infrared (NIR) pulses~\cite{schutte16a}. Amongst the various processes that take place during the cluster expansion, electron-ion recombination processes are known to play an important role~\cite{ditmire96,jungreuthmayer05,saalmann06,arbeiter11,ackad13,Arbeiter14}. Recently, recombination resulting in the production of a large number of excited atoms and ions was temporally resolved using pump-probe photoion and -electron spectroscopy~\cite{schutte14a,schutte14b,schutte15a}, showing similar dynamics for clusters ionized by intense XUV and intense NIR pulses, respectively. These results suggest that the charge distributions that can be measured after the nanoplasma has expanded may differ significantly from the transient charge distributions that exist prior to and during the expansion. Such a behavior is further supported by recent fluorescence spectroscopy experiments from mixed Ar-Xe clusters, where signatures of high Xe charge states were found~\cite{schroedter14}. Similarly, Iwayama \textit{et al.} reported a fluorescence signal from highly charged Ar ions following intense XUV ionization of Ar clusters~\cite{iwayama15}. In contrast, the average charge states observed in the final photoion spectra were much lower~\cite{hoener08, iwayama13} and suggested highly efficient recombination.

The investigation of nanoplasma relaxation dynamics has so far mainly focused on atomic clusters. One advantage of molecular clusters is that they consist of elements with different properties compared to noble gases. It was shown in a recent experiment on oxygen clusters~\cite{schutte15b} that autoionization (i.e. nonradiative decay) of multiply-excited atoms plays an important role following ionization by moderately intense NIR pulses. Similar processes were later found as well in the case of atomic clusters~\cite{schutte16b}. 

The study of charge recombination and nonradiative decay in extended systems is relevant for coherent diffractive imaging experiments in large (bio-)molecules~\cite{spence12}. In this context, the investigation of hydrocarbon clusters like methane~\cite{last02,hohenberger05,iwan12,dicintio13} and propane~\cite{toma02a,toma02b} clusters can be beneficial, as hydrogen and carbon are important constituents of organic molecules.  

Here, we study transient charges in expanding clusters following ionization of CH$_4$ clusters by intense NIR laser pulses. Even though the clusters become highly charged, we observe a strong suppression of the C$^{4+}$ ion signal that is attributed to efficient charge recombination processes leading to the formation of excited C$^{3+}$ ions. This interpretation is supported by pump-probe photoion and -electron spectroscopy experiments, where we demonstrate the presence in the nanoplasma of C$^{3+}$ ions in Rydberg states, which we attribute to recombination of C$^{4+}$ ions and electrons during the cluster disintegration. Furthermore, we find clear signatures of nonradiative decay processes in the expanding clusters, which are attributed to Auger decay. This observation shows that excited C$^+$ and C$^{2+}$ ions are transiently formed, which decay nonradiatively and are observed as C$^{2+}$ and C$^{3+}$ ions in the final ion charge spectrum.

The laser system and the experimental setup have been discussed before (see~\cite{schutte15b} for a figure of the experimental setup), and, therefore, only a brief description will be given here. We use a Ti:sapphire laser system at a central wavelength of 790~nm delivering pulses with energies up to 35~mJ and a minimum duration of 32~fs (full width at half maximum)~\cite{gademann11}. Within the amplifier unit, the laser beam is split into a pump and a probe beam. 2 independent grating compressors are used to control their pulse lengths. In the experiments, we use different pump pulse durations (40 and 400~fs), while the probe pulse duration is always 40~fs. Both compressed pulses are recombined by a mirror with a 6~mm central hole. While the pump beam is transmitted through the hole, the probe beam is reflected by the mirror. This - perhaps somewhat unusual - method for preparing the pump-probe optical path is a consequence of the fact that the experimental apparatus described in~\cite{schutte15b} is predominantly used for experiments involving XUV radiation formed by means of high-harmonic generation. The beams propagate collinearly towards a 75~mm focal length spherical mirror that focuses the pulses to the center of a velocity map imaging spectrometer~\cite{eppink97}. The laser pulses interact with a pulsed cluster beam that is generated by a piezoelectric valve with a nozzle diameter of 0.5~mm. A 0.2~mm diameter molecular beam skimmer selects the central part of the cluster beam. The average CH$_4$ cluster size is controlled by the backing pressure and is estimated as $\langle N \rangle$=15000 molecules according to the Hagena scaling law~\cite{hagena72}. In the experiment, ions and electrons resulting from the laser-cluster interaction are accelerated by a static electric field. The charged particles are detected with a multichannel plate / phosphor screen assembly, and projected 2D momentum maps are recorded with a CCD camera. 3D electron momentum distributions are obtained by an Abel inversion method~\cite{vrakking01}. Ion mass spectra are measured by using the velocity map imaging spectrometer in a time-of-flight (TOF) mode.


\begin{figure}[htb]
\centering
\includegraphics[width=6cm]{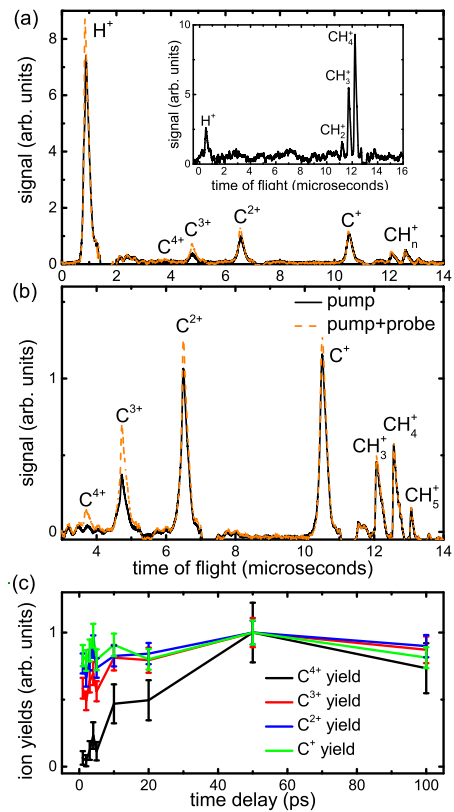}
\caption{\label{figure1} (a) Ion TOF spectrum resulting from the ionization of CH$_4$ clusters (solid black curve) and individual CH$_4$ molecules (inset) by a 40~fs NIR laser pulse ($I=8 \times 10^{13}$~W/cm$^2$). The orange curve shows the ion TOF spectrum when adding a second, weak NIR laser pulse ($I=5 \times 10^{12}$~W/cm$^2$) at a time delay of 50~ps with respect to the NIR pump pulse. We note that, inevitably, the cluster measurement contains contributions from the interaction of uncondensed molecules with the NIR pulse. (b) Zoom into the TOF spectrum from clusters that highlights the molecular and C$^{n+}$ fragments with and without the probe pulse. (c) Ion yields at different time delays between the NIR pump and probe pulses normalized to the maximum yields observed at 50~ps.
}
\end{figure}

Fig.~1(a) depicts an ion TOF spectrum recorded for CH$_4$ clusters that were ionized by an NIR pulse at an intensity of $8\times 10^{13}$~W/cm$^2$ and a pulse duration of 40~fs (solid black curve). Whereas TOF spectra for the ionization of isolated gas-phase CH$_4$ molecules using these laser conditions (see inset) are dominated by the presence of CH$_3^+$ and CH$_4^+$ ions, the TOF spectrum in the cluster experiment is dominated by the presence of H$^+$ and C$^{n+}$ ions. The differences between the final ion charge distributions observed from molecules and clusters are due to very efficient ionization avalanching in the latter case~\cite{ditmire96, schutte16a}. The multiphoton and / or tunneling ionization by the laser are complemented in the cluster by laser-driven electron impact ionization processes, using electrons that have been produced earlier in the pulse. This does not only result in a high ionization degree of the clusters as previously observed for atomic clusters~\cite{snyder96}, but also leads to a strong molecular fragmentation. We note that the degree of molecular fragmentation and the average ion charge state at these moderate NIR intensities are significantly higher than in a recent work on CH$_4$ clusters interacting with intense XUV pulses from a free-electron laser at a photon energy of 92~eV~\cite{iwan12}. Similar to that study, a CH$_5^+$ peak is visible in Fig.~1(b), which was previously explained by molecular recombination processes in the expanding cluster. This result is consistent with our understanding that while the ionization mechanisms are completely different for intense NIR and XUV laser pulses, qualitatively similar processes take place during the expansion and relaxation of clusters that have interacted with intense XUV and NIR laser pulses~\cite{schutte14a,schutte14b,schutte15a,schutte15c,oelze16}.

The formation of excited atoms and ions is probed by a second weak NIR pulse. Fig.~1(a),(b) shows ion TOF traces recorded with the additional probe pulse at a time delay of 50~ps (dashed orange curves), which leads to an enhancement in the yields of all atomic fragments. Since the NIR probe intensity of $5 \times 10^{12}$~W/cm$^2$ is not high enough to ionize atoms in the ground state, the enhancement is attributed to ionization of excited atoms and ions that are formed during the cluster expansion. Interestingly, a new contribution that is assigned to C$^{4+}$ ions emerges in the TOF spectrum. The observation of this peak points at the existence of high-lying C$^{3+}$ Rydberg states in the expanding nanoplasma. In line with our earlier work~\cite{schutte14b,schutte15a}, we attribute the formation of these Rydberg states to electron-ion recombination involving a transient C$^{4+}$ ion contribution. We note that electron impact excitation may also contribute to the formation of excited C$^{3+}$ ions. However, when these excited ions are formed during the laser-cluster interaction, ionization during the falling edge of the pump pulse (which is 16 times more intense than the probe pulse) leading to the formation of C$^{4+}$ ions is likely to occur. After the laser pulse has ended, rapid expansion cooling of the electrons takes place. It is therefore suggested that most excited atoms and ions are formed by charge recombination, which could be unambiguously demonstrated in the case of excited neutral atoms~\cite{schutte14b,schutte15a}.


As shown in Fig.~1(c), the C$^{4+}$ ion yield (black curve) increases with increasing delays between the 2 NIR laser pulses before saturation sets in at 50~ps. A much smaller increase is observed for the other charge states. The increasing yields can be influenced by (i) the time that it takes for ions and electrons to recombine, and (ii) by the possibility to probe this recombination using the probe pulse. Since the probe pulse generates electrons with kinetic energies that are typically $<1.6$~eV~\cite{schutte14b}, these electrons can only escape the cluster, when the potential is sufficiently shallow. As the cluster potential decreases during the nanoplasma expansion, the probability of electron emission induced by the probe pulse from the cluster increases.

\begin{figure}
\centering
\includegraphics[width=6cm]{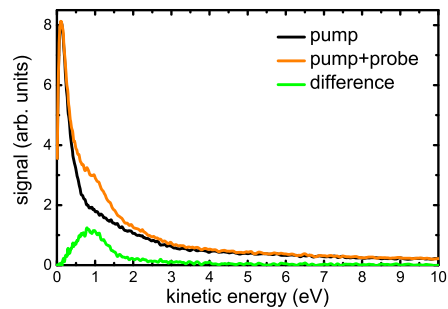}
\caption{\label{figure2} Electron kinetic energy spectra from CH$_4$ clusters, where the distributions for the pump pulse only, for pump+probe pulses at a time delay of 100~ps and the difference between these are shown. The NIR pump and probe pulses have intensities of $8 \times 10^{13}$~W/cm$^2$ and $5 \times 10^{12}$~W/cm$^2$, respectively.} 
\end{figure}

Indications of charge recombination are also observed in measured electron spectra from CH$_4$ clusters shown in Fig.~2, where the contributions with (orange curve) and without probe pulse (black curve) are shown. The difference spectrum (green curve) is composed of a broad contribution that peaks near 1~eV. Similar to our earlier studies in atomic clusters~\cite{schutte14b, schutte15a}, this contribution is attributed to the ionization of Rydberg atoms and ions with a binding energy up to 1.57~eV (the NIR photon energy). The signal at higher kinetic energies can be attributed to ionization processes using 2 or more NIR photons. We note that, unlike in~\cite{schutte14b}, individual peaks corresponding to specific excited states cannot be resolved in the electron spectrum due to the large number of states from different ionic fragments that can be involved. 
 																											
\begin{figure}
\centering
\includegraphics[width=6cm]{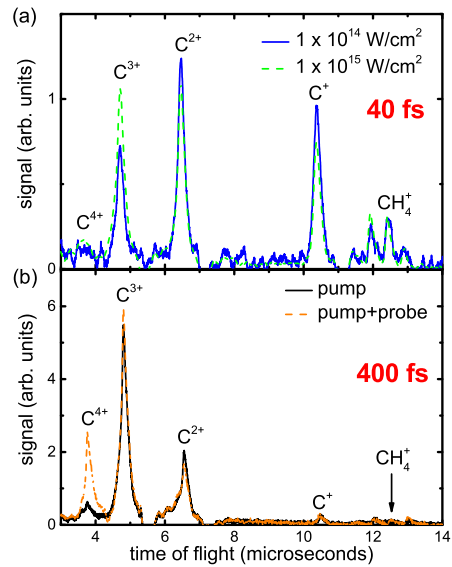}
\caption{\label{figure3} (a) Ion TOF spectra from CH$_4$ clusters at NIR intensities of $1 \times 10^{14}$~W/cm$^2$ and $1 \times 10^{15}$~W/cm$^2$, using a pulse length of 40~fs. (b) Ion TOF spectra for an increased pulse length of 400~fs ($I=1 \times 10^{14}$~W/cm$^2$). 2 spectra are shown with and without an NIR probe pulse ($I=5 \times 10^{12}$~W/cm$^2$) and a time delay of 50~ps between the pulses.}
\end{figure}

In order to further elucidate the processes associated with the dynamics of the C$^{4+}$ transient charge state, the pump laser conditions were varied. In Fig.~3(a), pump-only ion TOF spectra are shown for NIR intensities of $1 \times 10^{14}$~W/cm$^2$ (solid blue curve) and $1 \times 10^{15}$~W/cm$^2$ (dashed green curve). The average charge state slightly increases with intensity, but the C$^{4+}$ ion yield remains small at an intensity of $1 \times 10^{15}$~W/cm$^2$. This could be the result of a dominating signal from lower-intensity regions within the NIR focal volume.

As a next step, we have applied longer laser pulses, leading to more efficient ionization avalanching~\cite{schutte16a}. In a previous study on Xe clusters using NIR laser pulse durations between 175 and 1200~fs, a step-like behavior in the ion charge state distribution was observed, once the laser reached the threshold intensity for tunnel ionization~\cite{doppner10}. In this case, the ion charge state distributions only weakly depended on the NIR intensity. In Fig.~3(b), a pump-only ion TOF spectrum is shown for an NIR pulse duration of 400~fs and an intensity of $1 \times 10^{14}$~W/cm$^2$ (solid black curve). Compared to the result shown in Fig.~3(a), the charge distribution shifts to higher charges, with C$^{3+}$ dominating and C$^+$ being strongly reduced. However, like before, the C$^{4+}$ contribution remains very small. This is very surprising, because the third and fourth ionization potentials of C (47.9~eV and 64.5~eV) are rather similar. When switching on the probe laser pulse in Fig.~3(b) at a delay of 50~ps (dashed orange curve), the C$^{4+}$ ion contribution is strongly enhanced, whereas the C$^{2+}$ ion yield is slightly reduced. The latter observation is attributed to ionization of excited C$^{2+}$ ions by the probe pulse, which are then detected as C$^{3+}$ or C$^{4+}$ ions. This is in contrast to our previous investigation on NIR ionization of mixed Ar-Xe clusters~\cite{schutte15a}, where the yields of all charge states were increased when applying a weak NIR probe pulse. We conclude that the only way to observe substantial formation of C$^{4+}$ in our experiment is to arrange conditions where the formation of the detected C$^{4+}$ occurs after the cluster has had time to disintegrate. When C$^{4+}$ is formed by the pump pulse at times when the cluster is still intact, no C$^{4+}$ ions are detected, presumably because any C$^{4+}$ ions that are formed recombine with at least one electron, forming an excited ion or atom.

\begin{figure}
\centering
\includegraphics[width=6cm]{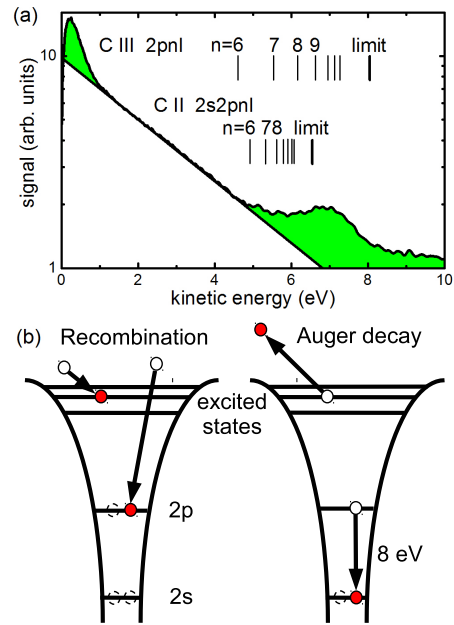}
\caption{\label{figure4} (a) Electron spectrum from CH$_4$ clusters ionized by 400~fs NIR pulses ($I=1 \times 10^{14}$~W/cm$^2$). In addition to an exponential contribution, a clear peak is observed at 7~eV, which is attributed to Auger decay of doubly-excited C$^{+}$ and C$^{2+}$ ions. The limit of the C II 2s2pnl series was taken from~\cite{moore70} and the limit of the C III 2pnl series is from~\cite{edlen83}. The lower excited states were calculated by a Rydberg formula in agreement with~\cite{schippers01}. We note that the yield of electrons with kinetic energies $< 1$~eV is probably underestimated due to saturation at the center of the detector. (b) Recombination processes in expanding clusters that lead to possible subsequent Auger decay, shown for the case where the recombination results in the formation of C III 2pnl. After recombination, the ion can relax to the ground state, releasing the second excited electron into the continuum.}
\end{figure}

In Fig.~4(a), an electron kinetic energy spectrum is shown following ionization of CH$_4$ clusters by 400~fs NIR pulses ($I=1 \times 10^{14}$~W/cm$^2$). The predominant exponential contribution (indicated by the straight line in the logarithmic plot) may be attributed to thermal and direct electron emission~\cite{ditmire96}. In addition, a peak is visible at a kinetic energy of 7~eV, which can be explained by the emission of electrons that are initially in bound atomic states. We assign this peak to nonradiative decay processes of C II 2s2pnl states (leading to the ejection of electrons with a maximum energy of 6.5~eV~\cite{moore70}) and C III 2pnl states (leading to the ejection of electrons with a maximum energy of 8.0~eV~\cite{edlen83}). This would result in the generation of C$^{2+}$ and C$^{3+}$ ions, which is consistent with the main contributions observed in the ion TOF spectrum shown in Fig.~3(b). Generally, the electron emission can be attributed to correlated electronic decay processes, where one electron relaxes from an excited to the ground state and transfers the excess energy to a second electron that leaves the cluster~\cite{schutte15c, oelze16}. Since in the discussed processes an inner-valence vacancy is filled, they can more specifically be described as Auger decay~\cite{auger25}, which, to our knowledge, has not been previously observed following ionization of clusters by intense NIR pulses. Our experiments demonstrate that nonradiative decay processes in expanding clusters are important even at very high ionization degrees.

The observed Auger decay can be explained by a 3-step process. In a first step, the NIR pump laser pulse removes almost all valence electrons from the C atoms. In a second step, as shown in Fig.~4(b), at least 2 electrons populate Rydberg and outer valence state levels. In a third step, one electron relaxes to a 2s vacancy and transfers the excess energy to a second electron that leaves the cluster. 

Previous studies have shown that only those nonradiative decay processes, which occur at least a few picoseconds after the laser pulse has ended, can be identified in electron kinetic energy spectra~\cite{schutte15b, schutte15c, oelze16, schutte16b}. At earlier times, the cluster potential strongly influences the kinetic energies of emitted electrons. Similarly, in the present experiment, only nonradiative decay processes occurring on picosecond to nanosecond timescales can be identified in the electron spectra. The timescale of electron emission is influenced both by the time it takes to form autoionizing states and by the decay time of these states. Note that such comparably slow Auger decay processes were predicted~\cite{safronova96,safronova97} and can partly be rationalized by the involvement of electrons in Rydberg states. In contrast, nonradiative decay processes taking place on a femtosecond timescale~\cite{safronova96,safronova97} do not leave clear signatures in the spectra, and should be the subject of future theoretical and experimental work. We note that autoionizing states leading to the emission of electrons with kinetic energies above $8$~eV~\cite{safronova96,safronova97} can be populated as well and may explain the signal exceeding the exponential curve in Fig.~4(a).

In summary, we have reported an investigation of transient ion charges in expanding CH$_4$ clusters. By using a weak NIR probe pulse that ionizes excited ions, we concluded that transient ion charges exist, whose charge states were \textit{decreased} by electron-ion recombination. We further found evidence for transient ion charges, whose charge states were \textit{increased} again by nonradiative decay processes. The observation of Auger decay demonstrates that electron correlation remains important in the relaxation dynamics of highly charged clusters and should be included in improved model calculations. The current results are expected to be generic for extended systems such as atomic clusters, nanoparticles and large molecules interacting with intense light pulses at different wavelengths, specifically including the XUV and x-ray regimes.

%
%


%

\end{document}